\documentclass[twocolumn,prl,showpacs,preprintnumbers,amsmath,amssymb,floatfix]{revtex4}
\usepackage{graphicx}
\usepackage{epsfig}

\begin{document}

\title{Resonant superradiant backward-scattering as a source for many-particle entanglement}

\author{K.M.R. van der Stam}
\affiliation{Atom Optics and Ultrafast Dynamics, Utrecht University,\\
P.O. Box 80,000, 3508 TA Utrecht, The Netherlands}
\author{R. Meppelink}
\affiliation{Atom Optics and Ultrafast Dynamics, Utrecht University,\\
P.O. Box 80,000, 3508 TA Utrecht, The Netherlands}
\author{J.M.~Vogels}
\affiliation{Atom Optics and Ultrafast Dynamics, Utrecht University,\\
P.O. Box 80,000, 3508 TA Utrecht, The Netherlands}
\author{J.W. Thomsen}
\altaffiliation[Permanent address: ]{Niels Bohr Institute, Universitetsparken 5, 2100 Copenhagen, Denmark}
\author{P. van der Straten}
\email{P.vanderStraten@phys.uu.nl}
\affiliation{Atom Optics and Ultrafast Dynamics, Utrecht University,\\
P.O. Box 80,000, 3508 TA Utrecht, The Netherlands}

\date{\today }

\begin{abstract} 
We investigate the atom pair production by superradiant backward-scattering from a Bose-Einstein condensate. By driving the superradiant process with two frequencies we can extend both the range of pulse duration and intensity by two orders of magnitude and obtain full control over the number of scattered atoms in forward and backward direction. We show that the atoms scattered in forward direction are strongly correlated with the atoms scattered in backward direction, which makes resonant superradiant backward-scattering a promising candidate for many-particle entanglement. 
\end{abstract} 

\maketitle

Superradiant emission is observed when a Bose-Einstein condensate (BEC) is illuminated with a single near-resonant laser beam. The superradiance is generated by collective Rayleigh scattering from the condensate, where the scattering rate is strongly enhanced compared to single atom scattering. A Rayleigh scattering event leads to a recoiling atom, which interferes with the atoms at rest. The interference pattern in the density distribution acts as grating, which stimulates the next atom to scatter in the same direction making the process self amplifying. In an elongated BEC the amplification is strongest along the long axis of the BEC due to the larger solid angle contributing to the same atom mode, the so-called endfire modes~\cite{Haroche}. In this process the atom will absorb a photon from the incoming laser beam followed by emission along the long axis of the BEC. When the incoming laser beam is orientated perpendicular to the long axis of the BEC, the recoil of the atom will be at an angle of $\pm$45$^\circ$ with respect to the incoming laser beam. This scattering process was first observed by Inouye \textit{et al.}~\cite{Ketterle1} and followed by various distinct theoretical descriptions~\cite{Muste,Meystre,Benedek,Zobay}.

A second scattering process has been identified when an atom absorbs a photon from the endfire beam followed by stimulated emission into the laser beam. The recoil of the atom will in that case be directed backwards at an angle of $\mp$135$^\circ$ with respect to the incoming laser beam \cite{Ketterle2}. This superradiant backward-scattering, which is referred to as Kapitza-Dirac scattering in Ref. \cite{Ketterle2}, is a promising way of generating many-particle entanglement \cite{Meystre}. The entanglement produced by superradiant backward-scattering results in physically separated ensembles containing a large number of massive particles. This provides a step forwards for the practical application of entanglement \cite{Polzik, Polzik2, Sorensen}.

After the two successive Rayleigh scattering events creating forward- and backward-scattering, two atoms move in opposite directions with each a velocity of $\sqrt{2}$ times the recoil velocity. Since the photon from the endfire mode is scattered back into the laser beam, this leads to an energy deficit of four times the recoil energy $E_r$. In Ref. \cite{Ketterle2} this lack of energy conservation is circumvented by using a pulse length, for which the energy uncertainty is larger than the energy mismatch. In the case of sodium the energy mismatch is 100 kHz resulting in a maximum pulse length of 10~$\mu$s leading for instance to a not well-defined phase of the light field, which limits the applicability of the process.

\begin{figure}[t!]
\centering
\includegraphics[width=0.35\textwidth]{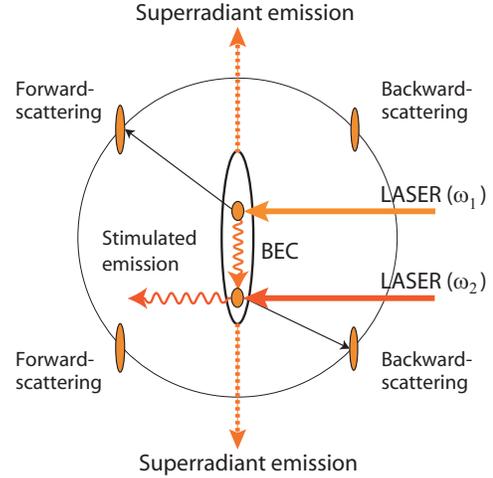}
\caption{Schematic representation of resonant backward-scattering. By absorption of a photon from the first laser beam ($\omega_1$) and emission in the endfire mode an atom is scattered forwards. The emitted photon is absorbed by another atom followed by stimulated emission into the second laser beam ($\omega_2$) resulting in backward-scattering. The frequency difference between the two lasers is used to drive this process resonantly. } \label{KDresonantekdopzet}
\end{figure}

In the experiments described in this Letter we will drive the backward-scattering resonantly by using two laser frequencies, which   yields a full control over the process (Fig. \ref{KDresonantekdopzet}). In this scheme we can transfer a large fraction of the atoms to the backward-scattered mode independent of the pulse length. The ratio between the number of forward- and backward-scattered atoms is controlled by the frequency difference between the two lasers. The experimental results show that the generation of atom pairs by resonant backward-scattering is preferred over single atom scattering even at low laser intensities, which makes it a robust process. Furthermore, the behavior of the process as a function of the pulse length shows a strong number correlation yielding a first indication that the system exhibits many-particle entanglement.

In our setup a sample of sodium atoms trapped in a cloverleaf magnetic trap is cooled by evaporative cooling to degeneracy in 50~s. The trap frequencies in the axial and radial direction are $\nu_{z}$ = 4~Hz and $\nu_{\rho}$ = 99~Hz, respectively. The BEC contains 1.2$\cdot$10$^{8}$ sodium atoms in the $F_g=1$, $M_g=-1$ ground state at a temperature of 450 nK and a density of 2.8$\cdot$10$^{14}$ atoms/cm$^{3}$ \cite{spinpol,grootsteBEC}. The axial and radial size of the condensate are $\ell$ = 1.0~mm and $d$ = 40~$\mu$m, respectively. The very elongated shape of the condensate results in a Fresnel number $F=\pi d^2/4\ell\lambda$ of 2 allowing for only a few dominant endfire modes in each direction along the long axis of the BEC \cite{Haroche}. 

To generate resonant backward-scattering the BEC is illuminated with two laser beams with joint detuning ranging from $-$5 GHz to +5 GHz from the $F_g=1$ to $F_e=2$ transition. The width of the laser beams is larger than the size of the BEC and therefore the profile of the beams can be assumed to be homogeneous over the BEC. The lasers are incident on the BEC under a small angle ($\approx$5$^\circ$) with respect to each other; however, using only one beam with two frequencies also worked. Light polarization of the two beams is perpendicular to the long axis of the BEC allowing for Rayleigh scattering in that direction. Time duration, intensity and detuning of the laser pulses as well as the frequency and phase difference between them are adjustable parameters in the experiment. After the illumination of the BEC the magnetic trap is switched off followed by a ballistic expansion of 30 ms. After the expansion an absorption image of the cloud is produced with a laser beam resonant to the $F_g=1$ to $F_e=1$ transition.

\begin{figure}[t!]
\centering
\includegraphics[width=0.49\textwidth]{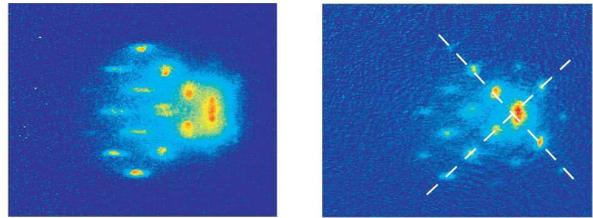}
\caption{The results for superradiant scattering using a pulse length of 100~$\mu$s, an intensity of 1.6~mW/cm$^2$ and a detuning of $-$3~GHz, where the largest cloud is the original condensate. The picture on the left shows the momentum distribution after illumination  by a laser beam incident from the right with one frequency. The picture on the right shows the results after illumination with two laser beams incident from the right with a frequency difference of 100 kHz. } \label{KDvb-resKD}
\end{figure}

In Fig. \ref{KDvb-resKD} results for superradiant scattering are shown. In the figure on the left the BEC is illuminated for 100~$\mu$s using a single laser beam. The pulse length is too long to generate non-resonant backward-scattering and in the experiment we only observe scattering forwards. Due to the large number of atoms in the condensate we observe scattering to higher order modes and this leads to  the well-known fan-pattern of superradiant forward-scattering. As can be clearly observed from the figure the atoms are scattered from the outer ends of the condensate, where the endfire modes have the largest intensity. Thus the size of the scattered modes are smaller than the condensate. In the figure on the right the pulse length and detuning are identical with the figure on the left, but the intensity is equally distributed over two beams with a frequency difference of 100~kHz. The result shows a clear presence of the backward-scattered modes. Note, that the backward-scattered atoms moving top right in the figure are on the same line as the forward-scattered atoms moving bottom left and thus that these two modes are generated by the same endfire mode~\cite{Zobay}. 

\begin{figure}[t!]
\centering
\includegraphics[width=0.49\textwidth]{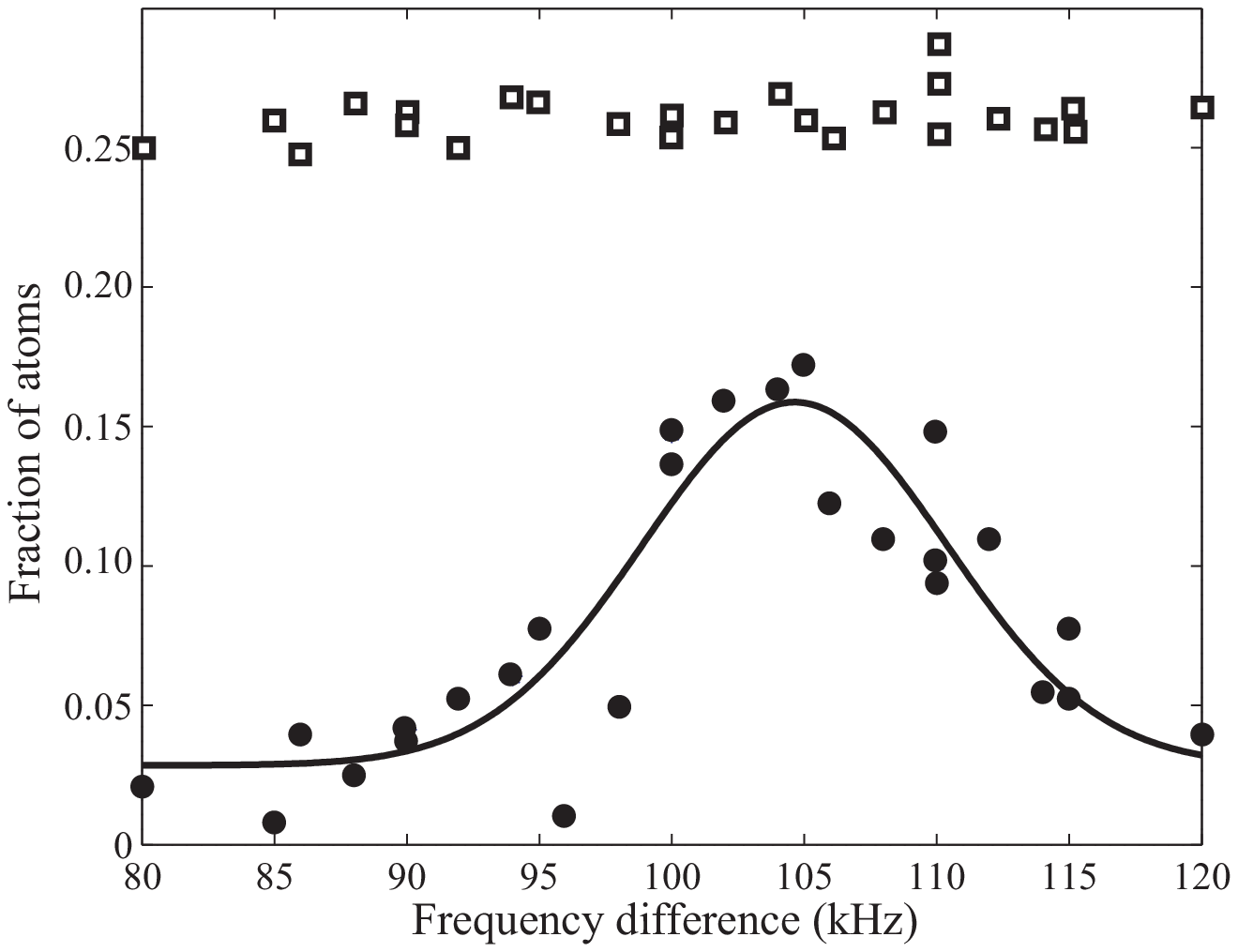}
\includegraphics[width=0.49\textwidth]{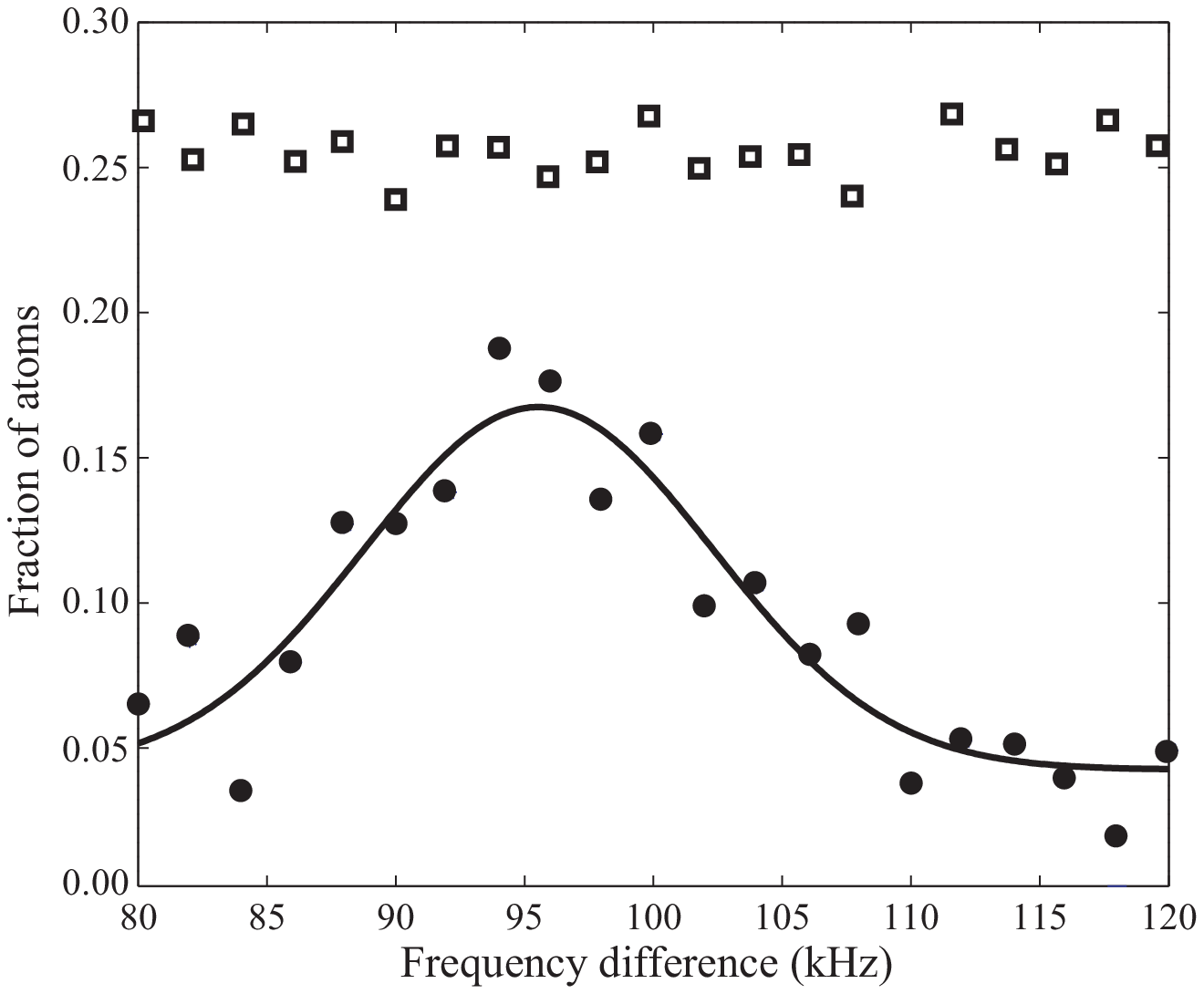}
\caption{The fraction of the atoms in the backward-scattered mode (circles) and the forward-scattered mode (squares) as a function of the frequency difference between the two laser beams. The upper picture is for a detuning of $-$5~GHz and a pulse length of 200~$\mu$s. The lower picture is for a detuning of +3.5~GHz and a pulse length of 100~$\mu s$. The intensity is 2.0 mW/cm$^2$ for both experiments. The solid lines are fits  to the data using a Gaussian function.} \label{KDResonantie}
\end{figure}

In Fig. \ref{KDResonantie} the fraction of atoms scattered forwards (squares) and backwards (circles) are shown as a function of the frequency difference between the two laser beams. Figure \ref{KDResonantie}a and b are for detunings of $-$5 GHz and +3.5 GHz, respectively. We observe a  resonance at  $104.5\pm0.6$ kHz for a detuning of $-$5 GHz and at $95.7\pm0.5$ kHz for a detuning of +3.5 GHz, which is clearly around $4 E_r/h$ = 100 kHz expected for sodium. The widths of the resonances are 17.6$\pm$1.6~kHz and 19$\pm$2~kHz. This broadening is caused by the mean-field of the condensate ($\mu$ = 5 kHz), the finite pulse duration and probably saturation.  In the second case the effective pulse length is shorter compared to the laser pulse length, since the endfire modes need a finite time to grow to their final strength. This limits the effective pulse duration to 50--100 $\mu$s, which is consistent with the measured width of the resonances.

The two resonances in Fig. \ref{KDResonantie} show a shift with respect to 100 kHz, where the shift has an opposite sign  compared to the sign of the detuning. The shift due to changing recoil frequency as measured by Campbell~\textit{et al.}~\cite{Ketterle3} is more than one order of magnitude too small and shows a different detuning dependence. The shift is probably caused by the light shift on the atoms induced by the endfire modes. For our geometry we find that at the end of the laser pulse the endfire modes reach an intensity level, which is close to the intensity of the incoming laser beams. The endfire modes are detuned by approximately the same detuning as the incoming lasers and the interference between the two endfire modes creates an optical lattice with a well-depth of a few kHz. Since the recoiling atoms have energies which are much larger than this well-depth, they will experience only an average light shift and thus will be shifted less compared to the condensate atoms. Note, that the condensed atoms in the lattice will also have higher order momenta components of $\pm 2 \hbar k$, which can be seen on the right in Fig. \ref{KDvb-resKD} in time-of-flight above and below the condensate. Validation of this assumption requires detailed knowledge about the strength and shape of the endfire mode, which goes beyond the scope of the present paper~\cite{future}.

The occupation of the forward-scattered mode shows no dependence on the frequency difference (see Fig. \ref{KDResonantie}), which shows that there is no suppression of the superradiant gain due to backward-scattering. This suppression has been predicted by Schneble \textit{et al.} \cite{Ketterle2} inspired by the idea that the gain in the endfire beam is attenuated due to the absorption of photons resulting in a suppression in the matter-wave amplification. In our experiment the endfire beam is largely absorbed in the condensate, since the number of forward- and backward-scattered atoms are almost equal. Although there is a strong attenuation of the endfire beam, no suppression of the forward-scattering is noticeable in Fig. \ref{KDResonantie}. We attribute this to the fact that in our experiment resonant atom-pair scattering is favored over single atom scattering. Note, that we can interpreted atom-pair scattering as a special case of four-wave mixing with matter waves~\cite{deng}, where in our case the light pulses supply the energy necessary for energy conservation.

\begin{figure}[t!]
\centering
\includegraphics[width=0.49\textwidth]{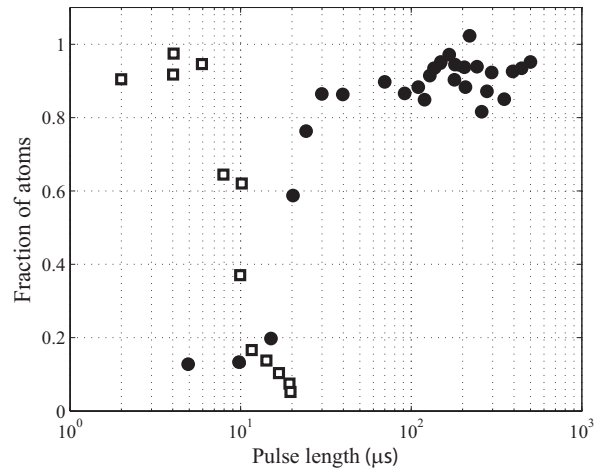}
\caption{The fraction of atoms in the backward-scattered mode normalized to the number of atoms in the forward-scattered mode as a function of the pulse length for non-resonant (squares) and resonant backward-scattering (circles). An intensity of 50~mW/cm$^2$ and 2 mW/cm$^2$ is used for the non-resonant and resonant case, respectively. The detuning is +3 GHz in both cases.} \label{KDKDvst}
\end{figure}

In Fig. \ref{KDKDvst} the fraction of atoms in the backward-scattered mode is measured as a function of the pulse length for non-resonant and resonant backward-scattering. For the non-resonant case the backward-scattering is only possible for very short pulse lengths. When the pulse length is longer than 6~$\mu$s the scattering process evolves from the short-pulse backward-scattering regime into the long-pulse forward-scattering regime, as indicated by the onset of an asymmetric distribution. This is very similar to the results presented in Ref. \cite{Ketterle2}. For the resonant backward-scattering an equal distribution of atoms over the backward- and forward-scattered modes is measured for pulse lengths up to 500 $\mu$s. The intensity in the latter case is reduced by a factor of 25 to avoid heating due to the long pulse duration. This shows that the required intensity for backward-scattering is relative low, which is in contradiction with the predictions of Ref \cite{Meystre}. The fact that the backward-scattering is not influenced significantly by decreasing the light intensity is another indication that the scattering in atom pairs is strongly preferred over single atom scattering, thereby making number squeezing more probable.

\begin{figure}[t!]
\centering
\includegraphics[width=0.49\textwidth]{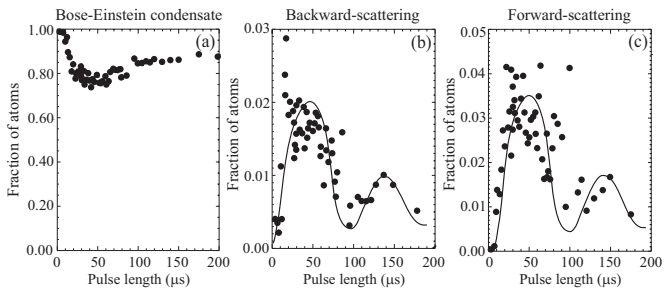}
\caption{The fraction of atoms  as a function of the pulse length normalized to the total number of atoms in (a) the BEC, (b) the backward-scattered mode and (c) the forward-scattered mode. The intensity is 4 mW/cm$^2$, the detuning is +3 GHz and the frequency difference is 100 kHz. The solid lines are guides to the eye. Note, that due to a smaller number of atoms in the condensate, the outcoupling fraction is smaller than the fraction shown in Fig.~\ref{KDResonantie}.} \label{KDrabiflopping}
\end{figure}

In Fig. \ref{KDrabiflopping} the fraction of atoms for forward- and backward-scattering is measured as a function of the pulse length. After an initial increase in both forward- and backward-scattering for short pulse durations, the fraction of atoms in both modes simultaneously decreases again. Most of these atoms are not scattered to higher order modes, but are scattered back into the condensate, as can be seen in  Fig. \ref{KDrabiflopping}a. This refilling of the condensate clearly shows that the coupling between the condensate and the two (forward and backward) modes occurs simultaneously in a coherent way and that the two modes remain coherent throughout the whole process. The decrease of the amplitude of the oscillation after pulse lengths of 180 $\mu$s  is due to scattering into higher-order modes and decoherence due to no longer overlap between the condensate and the scattered modes.

\begin{figure}[t!]
\begin{center}
\includegraphics[width=0.49\textwidth]{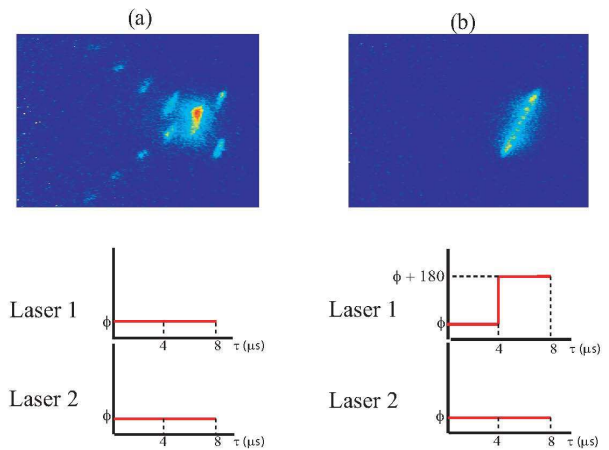}
\caption{(a) The scattering pattern after applying two laser pulses for 8 $\mu s$, where the phase $\phi$ of both lasers is kept constant during the process. (b) The scattering pattern for the same settings, but now the phase $\phi$ of one of the laser pulses is changed by 180$^\circ$ halfway during the pulse. The laser pulse has a detuning of +3 GHz and an intensity of 20 mW/cm$^2$.} \label{KDdraaiing}
\end{center}
\end{figure}

Another method of showing the correlation between the forward- and backward-scattered mode is refilling the BEC by reversing the phase of one of the laser beams halfway during the illumination. Figure \ref{KDdraaiing}a shows the results in case the phase between the two lasers is kept constant and we observe the X-shape pattern, which is characteristic for short-pulse backward-scattering. Figure \ref{KDdraaiing}b shows the result when halfway during the illumination the phase of one of the lasers makes a phase jump of 180$^\circ$. During the second half of the pulse the outcoupling process is completely reversed and this results in a complete refilling of the BEC. Attempts to refill the condensate using this method for longer pulses have been unsuccessful sofar. The interference pattern that can be observed in the condensate is probably caused by the motion of the atoms during the first half of the pulse, but its detailed nature is so far not completely understood. Such a reversal can only be achieved, if there is a strong correlation in the number of outcoupled particles as well as in their relative phases. Although this is not a sufficient criterion for entanglement, this is a strong indication that the system of resonant superradiant backward-scattering generates many-particle correlation in the system.

In this Letter we have shown that resonant backward-scattering provides control over the superradiant backward-scattering process. The atoms scattered in forward direction are strongly correlated to the atoms scattered in backward direction, which is shown by same temporal behavior of the outcoupling of forward and backward scattered atoms on resonance. In the case where the phase of one laser is reversed halfway through the pulse, the whole outcoupling of both forward and backward scattered atoms can be reversed. These results make resonant backward-scattering a promising technique to create many-particle entanglement.

We thank H.T.C. Stoof for fruitful discussions and for critical reading of the manuscript. This work is part of the research programme of the ``Stichting voor Fundamenteel Onderzoek der Materie'' (FOM), which is financially supported by the ``Nederlandse Organisatie voor Wetenschappelijk Onderzoek'' (NWO).

\end{document}